\RequirePackage{lineno}
\documentclass[aps,prb,showpacs, groupedaddress,twocolumn,superscriptaddress]{revtex4}
\usepackage{amsmath,graphicx,latexsym,times,color}
\usepackage{setspace}
\usepackage{hyperref}
\usepackage{array}
\usepackage{titlesec}
\usepackage{physics}

\begin{document}

\title{Giant anisotropic magnetoresistance through a  tilted molecular $\pi$-orbital}

\author{Dongzhe Li}
\email{donli@dtu.dk}
\affiliation{Department of Physics, University of Konstanz, 78457 Konstanz, Germany}
\affiliation{Department of Physics, Technical University of Denmark, DK-2800 Kongens Lyngby, Denmark}

\author{Fabian Pauly}
\affiliation{Okinawa Institute of Science and Technology Graduate University, Onna-son, Okinawa 904-0495, Japan}
\affiliation{Department of Physics, University of Konstanz, 78457 Konstanz, Germany}

\author{Alexander Smogunov}
\affiliation{SPEC, CEA, CNRS, Universit\'e Paris-Saclay, CEA Saclay 91191 Gif-sur-Yvette Cedex, France}

\date{\today}

\begin{abstract}
Anisotropic magnetoresistance (AMR), originating from spin-orbit coupling
(SOC), is the sensitivity of the electrical resistance in magnetic systems
to the direction of spin magnetization. Although this phenomenon has been
experimentally reported for several nanoscale junctions, a clear
understanding of the physical mechanism behind it is still elusive. Here we
discuss a novel concept based on orbital symmetry considerations to attain a
significant AMR of up to 95\% for a broad class of $\pi$-type molecular
spin-valves. It is illustrated at the benzene-dithiolate molecule connected
between two monoatomic nickel electrodes. We find that SOC opens, via
spin-flip events at the ferromagnet-molecule interface, a new conduction
channel, which is fully blocked by symmetry without SOC. Importantly, the
interplay between main and new transport channels turns out to depend
strongly on the magnetization direction in the nickel electrodes due to the 
tilting of molecular orbital. Moreover,
due to multi-band quantum interference, appearing at the band edge of nickel
electrodes, a transmission drop is observed just above the Fermi
energy. Altogether, these effects lead to a significant AMR around the Fermi
level, which even changes a sign. Our theoretical understanding, corroborated
in terms of \textit{ab initio} calculations and simplified analytical models, reveals
the general principles for an efficient realization of AMR in molecule-based
spintronic devices.
    
\end{abstract}

\renewcommand{\vec}[1]{\mathbf{#1}}

\maketitle

Spin-orbit coupling (SOC) is the quantum effect of relativistic nature, which links
electronic spin and orbital degrees of freedom. It is at the origin of a wide
range of intriguing phenomena in condensed matter physics such as the Rashba
effect, magneto-crystalline anisotropy, AMR, etc. Although AMR is the oldest
known magneto-transport effect, it is of high timeliness due to the recent
development of precise experimental tools to study magnetic systems at the
atomic scale. For instance, the tunneling AMR (TAMR) was first observed by Bode \textit{et al.}\cite{Bode2002} in scanning tunneling spectroscopy (STM), which was also reported in Ref.~\onlinecite{Bergmann2011,Ruppelt2013,schoneberg2018}. In addition, a very large TAMR was reported in various magnetic tunnel junctions \cite{Gould2004,Gould2005,Khalid2005,Burton2016}. In the contact regime, an enhanced ballistic AMR in atomic contacts was predicted theoretically \cite{Sabirianov2005,Jacob2008,Gabriel2008,Manrique2009,Viljas2009} and observed experimentally via mechanically controllable break junctions (MCBJ) \cite{Viret2006, Andrei2007,Bacca2010} or STM \cite{Otte2016-2}. Besides, an electrically tunable AMR was found in the Coulomb blockade regime in a ferromagnetic semiconductor single-electron transistor \cite{Jungwirth2006}. Recently, several experimental works \cite{Ryo2011,Wahler2011,li2015giant,rakhmilevitch2016,schoneberg2018,yang2019tunable}
on tuning AMR in single-molecule junctions have stimulated a new research
venue in molecular spintronics, which is the so-called molecular anisotropic
magnetoresistance (AMR) \cite{Otte2015}. Quantitatively, AMR is defined as
$\text{AMR} = (G_\parallel - G_\perp)/G_\perp$, where $G_\parallel$ and
$G_\perp$ are electrical conductances for parallel and perpendicular
orientations of the magnetization, respectively, with regard to the current
flow.

In bulk ferromagnetic metals, the AMR is less than 5\% \cite{mcguire1975} due
to quenched orbital moments. Its value can increase dramatically in
low-dimensional nanostructures such as monoatomic wires due to enhanced
orbital moments and the high sensitivity of the local electronic structure to
the magnetization direction induced by the SOC
\cite{gambardella2002,Jacob2008,Viljas2009}. Unlike metallic atomic contacts,
in molecular junctions the transport between two electrodes is typically
mediated by a relatively weakly bound molecule. Therefore, molecular orbitals
are expected to preserve their symmetry and localized nature. In the collinear
magnetic case without SOC, it has been shown that non-magnetic organic molecule can act as a ``half-metallic'' conductor due to either orbital symmetry arguments
\cite{smogunov2015,Dongzhe2019-Ni} or quantum interference effects
\cite{Li-QI-2019,pal2019}, leading to nearly fully spin-polarized conduction. In addition, a nearly perfect spin filtering was also reported when the vanadium-benzene wire is placed between two magnetic electrodes \cite{Maslyuk2006}. In the non-collinear magnetic case with SOC, both the band structures of the ferromagnetic electrodes and the selective hybridization between electrode and molecular states can be largely modified, yielding a large AMR.

Here, using fully relativistic density functional theory (DFT) calculations
combined with a scattering theory, we demonstrate how a giant
AMR of around 95\% at the Fermi level ($E_F$) can be obtained by designing a
molecular junction, in which molecular $\pi$ orbitals selectively hybridize
with $d$ bands of ferromagnetic electrodes. We discuss the mechanism using a
simple model system, consisting of a benzene-dithiolate (BDT) molecule joining
two semi-infinite monoatomic Ni chains. The conductance is fully
spin-polarized without SOC since the spin-up channel is blocked at the
ferromagnet-molecule interface by orbital symmetry mismatch between molecular
and electrode states. If SOC is switched on, a new spin-up-derived channel fully opens due to
spin-flip events. Furthermore, the SOC distinguishes the transmissions for
different magnetic orientations when the $\pi$-shaped molecular orbitals
couple to the Ni $d$ bands, giving rise to a very high and energy-dependent
AMR in the vicinity of $E_F$. We rationalize the DFT results by a simple
tight-binding model. Our findings provide guidelines of how an optimal AMR can
be achieved in $\pi$-conjugated molecular junctions based on clear symmetry
arguments.

\begin{figure}[!t]
	\centering
	\includegraphics[scale=0.41]{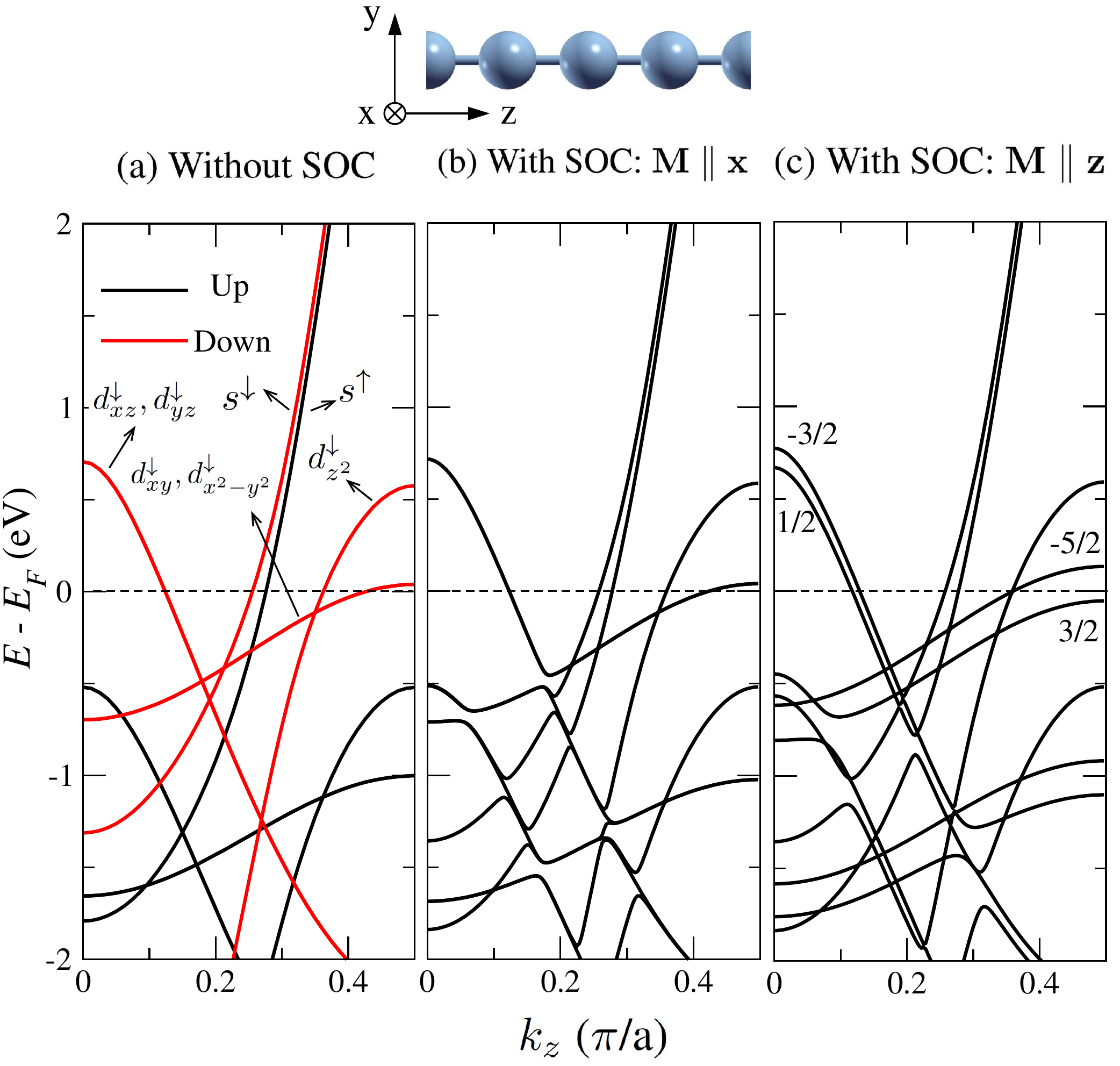}
	\caption{\label{ni-band} Band structure of a Ni monoatomic wire at the
		equilibrium lattice spacing of a = 2.10~\AA~without SOC (a) and with SOC for spin magnetization along the
		$\vec{x}$ (b) or the $\vec{z}$ (c) axis. The spin-up and -down bands
		in (a) are visualized by black and red lines, respectively. The bands are
		labeled by their orbital moment (a) or by total angular moment (c) along the
		wire axis $\vec{z}$.}
\end{figure}

The DFT calculations in the local density approximation (LDA) have
been performed using the plane-wave $\textsc{Quantum Espresso}$ (QE) package
\cite{Giannozzi2009}. The coherent electron transport was evaluated by
$\textsc{pwcond}$ \cite{Alexander2004} based on the scattering formalism
with ultra-soft pseudopotentials, which is a part of the QE package. The SOC, crucial for AMR, is taken into account via fully relativistic pseudopotentials \cite{Andrea2006}. The elastic
conductance is evaluated from the total electron transmission at the Fermi
energy using the Landauer-B\"uttiker formula, $G=G_0T(E_F)$, where
$G_0=e^2/h$ is the conductance quantum per spin. For collinear magnetic
systems without SOC effects the total transmission is the sum of two
independent spin transmissions,
$T(E_F)=T_\uparrow(E_F)+T_\downarrow(E_F)$. Structural optimizations of
molecular junctions have been performed without SOC using face-centered
cubic Ni(111) crystalline electrodes. For transport calculations they were
replaced with semi-infinite Ni chains. More details regarding computational
details can be found in the supplementary material.

\begin{figure}[htbp]
	\centering
	\includegraphics[scale=0.51]{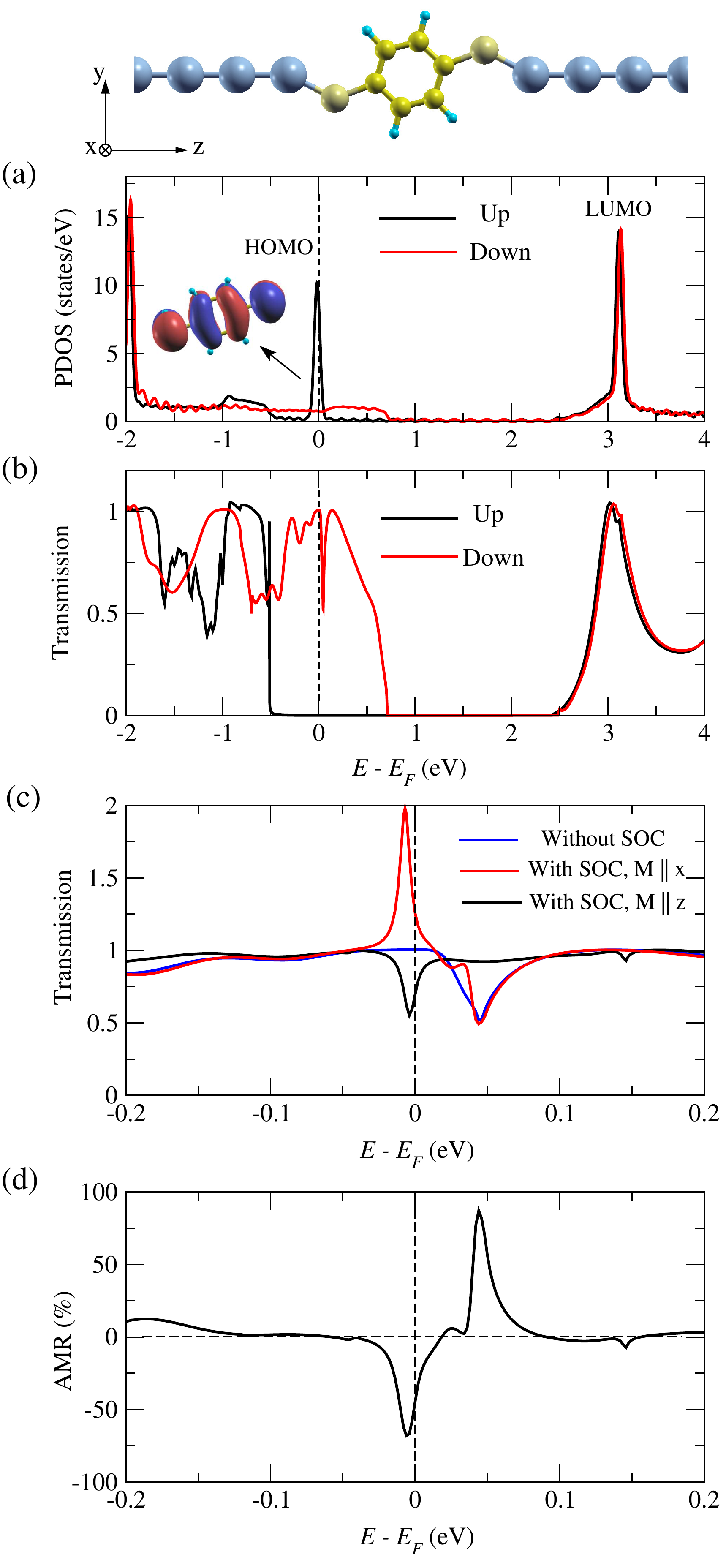}
	\caption{\label{dft-trans} 
		Ni wire/BDT junction:     
		(a) Spin-resolved PDOS on
		the BDT molecule without SOC. Wave function of the HOMO orbital of the free
		molecule is shown on inset. (b) Spin-resolved transmission function without SOC. A
		dip is observed for spin down right above the Fermi level. (c)
		Total transmissions of the junction without SOC (blue) and with SOC for $\vec{M}
		\parallel \vec{x}$ (red) and $\vec{M}\parallel\vec{z}$ (black) configurations. 
		(d) Energy-dependent AMR defined as
		$\text{AMR} (E) = (T_z (E) - T_x (E)/T_x (E)$. A giant AMR with a changing sign is found
		around $E_F$.}
\end{figure}

The objective of this work is to demonstrate theoretically the mechanism to
obtain giant molecular AMR effects based on orbital symmetry arguments. Here, we focus
on the influence of SOC on quantum transport across a BDT molecule sandwiched
between two semi-infinite Ni leads, as sketched in
Fig.~\ref{dft-trans}(a). Note that for better comparison to experiment electrodes with a larger cross section should be used, but we expect that our simplified model captures the relevant mechanisms, allowing at the same time a detailed analysis at reduced computational cost.

We start by studying the band structure of a Ni atomic chain, since it
provides information on the number of conduction channels in the
electrodes.  Let us first discuss the band structure of the Ni chain without
SOC, as plotted in Fig.~\ref{ni-band}(a). For spin-up (majority spin), only
one largely dispersive $s$ band crosses $E_F$ in the middle of the one-dimensional Brillouin zone, while six $d$ channels are available for spin-down (minority
spin). We mark explicitly two twofold degenerate bands, namely
$d_{xz}^{\downarrow},d_{yz}^{\downarrow}$ with a wide negative dispersion and
$d^{\downarrow}_{x^2-y^2},d^{\downarrow}_{xy}$ with a narrow positive
dispersion, which will be important in the following. When the SOC is
included, the band structures for magnetization $\vec{M}$ chosen parallel to
the $\vec{x}$ axis ($\vec{M} \parallel \vec{x}$) and $\vec{M} \parallel
\vec{z}$ are very different, as visible in Fig.~\ref{ni-band}(b) and
\ref{ni-band}(c). For $\vec{M}\parallel\vec{x}$ the band splitting by SOC is
tiny, so band dispersions are very similar to those without
SOC. Interestingly, a pseudo-gap opens at about $-0.45$~eV, finally causing a
large AMR of more than 160\% for the perfect Ni chain in that energy
region. For $\vec{M} \parallel \vec{z}$ the SOC lifts the degeneracy of both
$d^{\downarrow}_{xz},d^{\downarrow}_{yz}$ and
$d^{\downarrow}_{x^2-y^2},d^{\downarrow}_{xy}$ bands, resulting in sets of
$m_j=-3/2,1/2$ and $m_j=-5/2,3/2$ bands with similar dispersion, respectively,
where $m_j$ is the projection of the total angular momentum along the
$\vec{z}$ axis. The findings are in excellent agreement with previous
theoretical calculations \cite{Sabirianov2005,Otte2015}.

Now we discuss the transport properties of the Ni-BDT-Ni molecular junction,
shown in Fig. \ref{dft-trans}.  After geometry optimization with Ni(111)
crystalline electrodes, we find that the BDT molecule prefers to slightly
rotate in the $yz$ plane, which is consistent with previous theoretical
results \cite{Dmitry-2013,Karimi2016,Li_linking2019}. The highest occupied molecular
orbital (HOMO) of the BDT molecule is of odd symmetry with respect to the $yz$ plane, originating mainly from $p_x$ atomic orbitals of carbon and sulfur atoms. This is visible
in Fig.~\ref{dft-trans}(a), where the HOMO is presented together
with the projected density of states (PDOS) of BDT in the molecular junction
configuration. By symmetry, HOMO can only couple to $d_{xz}^{\downarrow}$ and $d_{xy}^{\downarrow}$ but not to $s$ states of the Ni chains. The PDOS therefore shows a very sharp HOMO peak for spin up around $E_F$ but a much broader feature for spin down due to larger hybridization. This is further reflected, see Fig.~\ref{dft-trans}(b), in a
complete blocking of the spin-up transmission around $E_{{F}}$, where only the
Ni $s$ channel is present, while a finite spin-down transmission is provided
by the Ni $d_{xz}^{\downarrow}$ channel, which generates a fully
spin-polarized conductance due to symmetry arguments proposed by us recently
\cite{smogunov2015}. Interestingly, a pronounced dip in the spin-down
transmission
is observed very close to the Fermi energy. It  
appears right above the Ni $d_{xy}^{\downarrow}$ band (see Fig.~\ref{ni-band}(a)) and 
results from destructive interference, as will be discussed later.

We now turn our attention to SOC effects on electron transport for different
spin magnetization configurations, namely $\vec{M} \parallel \vec{x}$ and
$\vec{M} \parallel \vec{z}$, as shown in Fig.~\ref{dft-trans}(c). We focus on
the energy range close to the Fermi energy.  Clearly, the transmission
functions for the two magnetic orientations are very different. In particular,
the transmission for $\vec{M} \parallel \vec{x}$ increases up to about 2 at about 20 meV below $E_{{F}}$, with $G_0= e^2/h$. This sharp peak originates
from the HOMO spin-up molecular orbital, which was inactive by symmetry
before, but couples with electrode states through the SOC term. For the
$\vec{M} \parallel \vec{z}$ configuration, on the contrary, a dip rather than
a peak is observed in the transmission at this energy, which results in an
$\text{AMR}$ as large as $-74\%$, see Fig.~\ref{dft-trans}(d).

\begin{figure}[!t]
	\centering
	\includegraphics[scale=0.42]{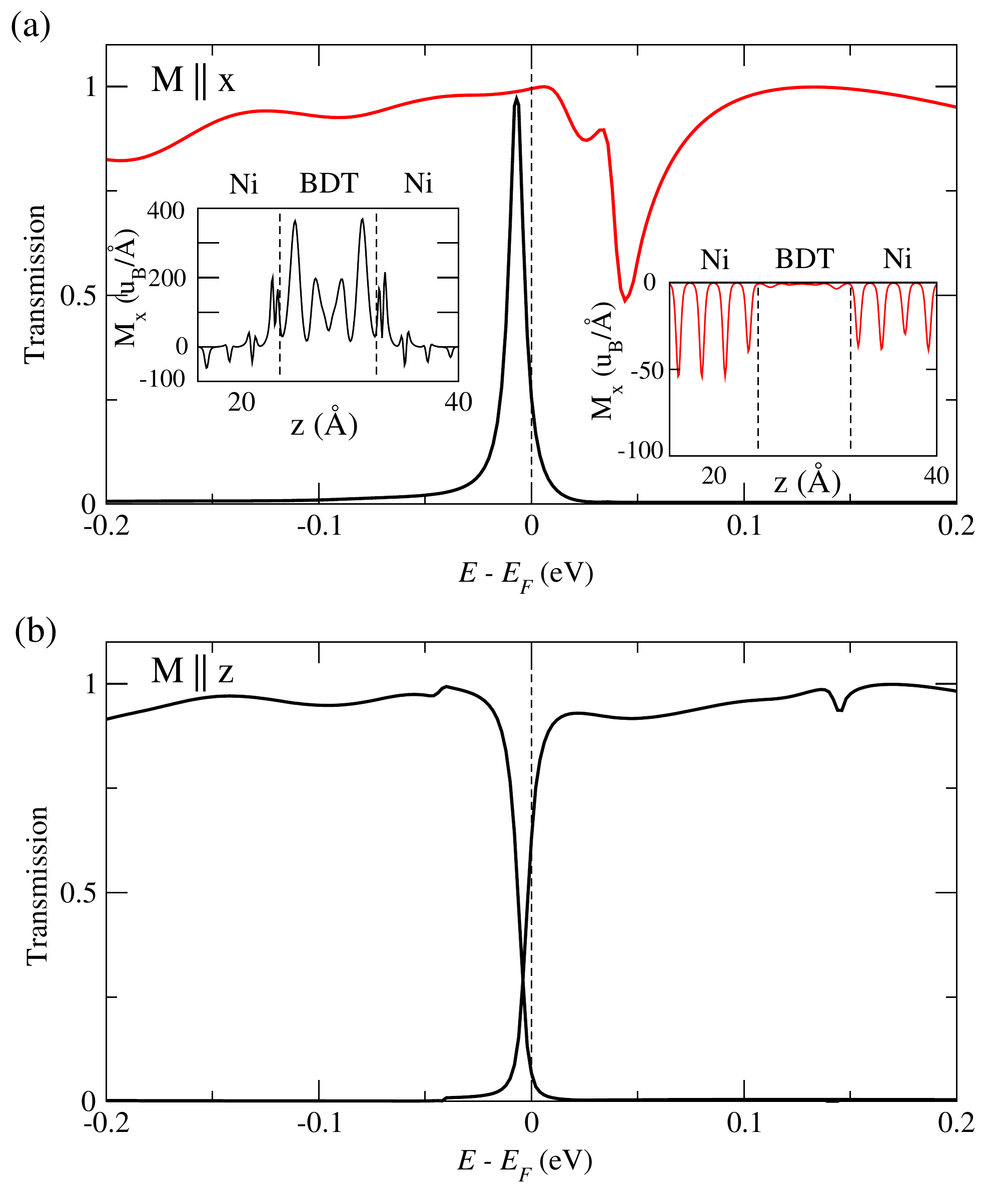}
	\caption{\label{T-eigenvalues} 
		{
			(a) Transmission eigenvalues for the $\vec{M}
			\parallel \vec{x}$ magnetic configuration, showing two independent
			channels (red and blue lines). Insets: magnetic moments (x-components), averaged in the $xy$ plane, as a
			function of $z$ for both eigenchannels at $E-E_F=-20$~meV. 
			Spin-flip (black) or spin-conserving (red) propagation of an electron is clearly seen for the HOMO-up or -down related channels, respectively. (b) Transmission eigenvalues for the $\vec{M} \parallel \vec{z}$ magnetic configuration, showing the mixing of two channels.
		}
	}
\end{figure}

To explore the origin of this huge AMR, we plot in Fig.~\ref{T-eigenvalues}
the transmission eigenvalues for both magnetic orientations. Two eigenchannels
are found for both cases instead of one spin-down eigenchannel without SOC.
The results for the $\vec{M} \parallel \vec{x}$ configuration indicate that
the two channels are independent. The $\vec{x}$ component of the magnetic moment ($\vec{M_x}$, averaged in $xy$ plane) of each channel at $E-E_F=-20$~meV, shown as insets, confirm that the highly transmissive
channel (red) is related to the HOMO spin-down orbital due to a slightly negative
spin moment on the molecule, while
the other one (black) stems from the HOMO spin-up orbital, as indicated by the
very large and positive spin moment on the BDT. A
similar conclusion is reached by comparing the eigenchannel transmissions for
$\vec{M} \parallel \vec{x}$ to the spin-resolved transmission in
Fig.~\ref{dft-trans}(b). Note that both channels describe the propagation of
electrons between spin-down Ni states: The first one (red) conserves the electron
spin while the second one (black) involves spin-flip processes at the metal-molecule
interfaces activated by the SOC term in the Hamiltonian.  For $\vec{M}
\parallel \vec{z}$, on the contrary, the two channels mix, exhibiting in
particular a crossing at $E-E_F=-20$~meV and a lower total transmission at that
$E$ compared to the $\vec{M} \parallel \vec{x}$ case, see
Fig.~\ref{dft-trans}(c).

Interestingly, the transmission for the $\vec{M} \parallel \vec{x}$
configuration shows a dip at about 40~meV above the Fermi energy (see
Fig.~\ref{dft-trans}(c)). It is again related to the edge of the Ni
$d^\downarrow_{x^2-y^2},d^\downarrow_{xy}$ bands, which is not modified by the
SOC in this situation. On the contrary, for $\vec{M} \parallel \vec{z}$ this
band is largely split into $m_j=3/2$ and $m_j=-5/2$ subbands (see
Fig.~\ref{ni-band}(c)). Consequently, the transmission dip moves with the
$m_j=-5/2$ band to higher energies and appears much less pronounced at around
150~meV. In summary, a very large AMR of a variable sign is found in the energy window between $-20$ and $60$~meV as demonstrated in Fig.~\ref{dft-trans}(d). Note that a large AMR of around 30\% has previously been measured for Ni-BDT-Ni molecular junctions by Yamada
\textit{et al.}~\cite{Ryo2011} and may be explained by our results.

Note that a range of similar metal-benzene complexes, including a model Ni/benzene junction, has been reported by Otte \textit{et al.}\cite{Otte2015}. The benzene molecule (without liking group of sulfur) was however oriented perpendicular to the transport direction (in the $xy$ plane) which is different to our geometry (Fig.~\ref{dft-trans}(a)). 
Very large AMR of about a few hundred were reported at $E-E_F=-450$ meV, attributed to the SOC-induced pseudo-gap in the Ni wire for $\vec{M} \parallel \vec{x}$ and to the orbital-symmetry filtering of the molecule at that specific energy (turning the pseudo-gap into the true transport gap). However, almost no AMR was found around the Fermi energy. Giant AMR ratios, found in our Ni/BDT juunctions close to the Fermi level, are generated, on the contrary, by SOC and interference effects at the molecule/Ni wire interfaces and do not rely on fine details of the Ni wire band structure. 
In particular, spin-flip processes at the Ni/molecule contacts open a new conduction channel (fully closed by symmetry in the absence of SOC), the interplay of which with another channels depends strongly  on the magnetization direction. 
We argue, therefore, that the physical mechanism behind giant AMR in our case is not the same as in Ref. [\onlinecite{Otte2015}]. 
Besides, we also reproduce extremely large AMR of more than 8000\% at $E-E_F=-450$ meV (see Fig.~S1 in the supplementary material).

Let us note that DFT contains uncertainties with regard to the energetic ordering of molecular levels and their alignment with the electrode states, while we expect the metallic Ni states to be well described. For improvements
computationally demanding quasiparticle methods, such as the GW approach,
would need to be coupled to our quantum transport calculations
\cite{Strange2011}. It may be argued that the quasiparticle corrections will
mostly affect the unoccupied orbitals by opening the HOMO-LUMO gap, while the
energy position of the HOMO is only slightly altered. Since we find that the
spin transport through the Ni-BDT-Ni system is dominated by the HOMO, we hope
that our DFT results and predictions are reliable. 

In order to explain our results for the AMR, we study the SOC term of the
Hamiltonian, which can be written as $H_{\text{SOC}}=\xi \vec{L}\cdot\vec{S}$,
where $\xi$ is the effective SOC constant and $\vec{L}$ and
$\vec{S}=\boldsymbol{\sigma}/2$ are the orbital momentum and spin operators of
an electron, respectively. In the following we will always fix the angular
momentum axis to the $\vec{z}$ direction, while we will choose the
spin-quantization axis along $\vec{x}$ or $\vec{z}$ for $\vec{M} \parallel
\vec{x}$ or $\vec{M} \parallel \vec{z}$ magnetic configurations,
respectively. For $\vec{M} \parallel \vec{x}$ the effective spin-orbit
Hamiltonian can thus be written as a $2 \times 2$ matrix in spin space,
\begin{equation}\label{SOC-X}
H_{\text{SOC}}^{x}={\xi \over 4}
\begin{bmatrix} 
{L}_{+}+{L}_{-} & -i(L_+-L_-)-2iL_z    \\
-i(L_+-L_-)+2iL_z & -{L}_{+}-{L}_{-} \\
\end{bmatrix}, 
\end{equation}
where $L_{\pm}=L_x \pm iL_y$. For $\vec{M} \parallel \vec{z}$ the same SOC
Hamiltonian has the form
\begin{equation}\label{SOC-Z}
H_{\text{SOC}}^z={\xi \over 2}
\begin{bmatrix} 
L_{z} & L_{-}    \\
L_{+} & -L_{z}   \\
\end{bmatrix}.
\end{equation}

As discussed before, by symmetry the molecular HOMO can only hybridize with
$d_{xz}$ and $d_{xy}$ Ni orbitals. Therefore, on Ni apex atoms, where the SOC is
essential, it can be expressed as	
\begin{equation}\label{homo-wave}
\begin{aligned}
\ket{\Psi_{\text{HOMO}}^\alpha}=A_\alpha\ket{d_{xz}}+B_\alpha\ket{d_{xy}} \\
=A_\alpha(\ket{-1}-\ket{1})+B_\alpha(\ket{-2}-\ket{2})
\end{aligned}
\end{equation}
with $\alpha=\uparrow,\downarrow$ and some spin-dependent coefficients
$A_\alpha$ and $B_\alpha$.  Here, the real harmonics $d_{xz}$ and $d_{xy}$
(not necessarily normalized) 
are expanded in terms of complex ones with orbital moment $m = \pm1$ and $m = \pm2$.

In the absence of SOC the HOMO spin-up orbital is decoupled from the Ni
electrodes, where only the $s$-band is available around the Fermi energy. We
apply now the SOC Hamiltonian to the HOMO spin-up orbital at the Ni apex atoms. For
the $\vec{M} \parallel \vec{x}$ orientation we get
\begin{small}
	\begin{equation}\label{x-homo}
	\begin{aligned}
	H_{\text{SOC}}^x\ket{\Psi^\uparrow_{\text{HOMO}}}=\\{\xi \over 2}
	\begin{bmatrix} 
	B_\uparrow\ket{d_{xz}}+A_\uparrow\ket{d_{xy}}     \\      -\sqrt{6}iA_\uparrow\ket{0}-i(B_\uparrow+A_\uparrow)\ket{d_{yz}}+i(A_\uparrow-2B_\uparrow)\ket{d_{x^2-y^2}} \\
	\end{bmatrix}
	\end{aligned}
	\end{equation}
\end{small}

We notice that a nonzero spin-down component will couple the HOMO spin-up orbital to spin-down Ni bands of mainly $d_{yz}^\downarrow, d_{x^2-y^2}^\downarrow$ character available at $E_F$. So a new
conduction channel will be opened through spin-flip processes. On the other
hand, the HOMO spin-down orbital will mainly conduct through the other
spin-down Ni bands of $d_{xz}^\downarrow,d_{xy}^\downarrow$ character. Moreover, it is clear that $\bra{\Psi^\downarrow_{\rm HOMO}}H_{\rm SOC}^x
\ket{\Psi^\uparrow_{\rm HOMO}}=0$, so that HOMO spin-up and spin-down orbitals
remain strictly orthogonal, rendering the two conduction channels
independent. For the $\vec{M} \parallel \vec{z}$ orientation an analogous
reasoning leads to
\begin{equation}\label{z-homo}
H_{\text{SOC}}^{z}\ket{\Psi^\uparrow_{\text{HOMO}}}={\xi \over 2}
\begin{bmatrix} 
-A_\uparrow\ket{d_{yz}}-2B_\uparrow\ket{d_{x^2-y^2}}     \\
A_\uparrow(\sqrt{6}\ket{0}-2\ket{2})+2B_\uparrow\ket{-1} \\
\end{bmatrix}.
\end{equation}
As before a nonzero spin-down component will open a new HOMO spin-up-related
channel, making two channels available for transport. One can observe however
that now $\bra{\Psi^\downarrow_{\rm HOMO}}H_{\rm SOC}^z
\ket{\Psi^\uparrow_{\rm HOMO}} \ne 0$. For this reason the two channels will
be mixed. The conclusions confirm our previous observations in the context
of Fig.~\ref{T-eigenvalues}.

\begin{figure}[!t]
	\centering
	\includegraphics[scale=0.51]{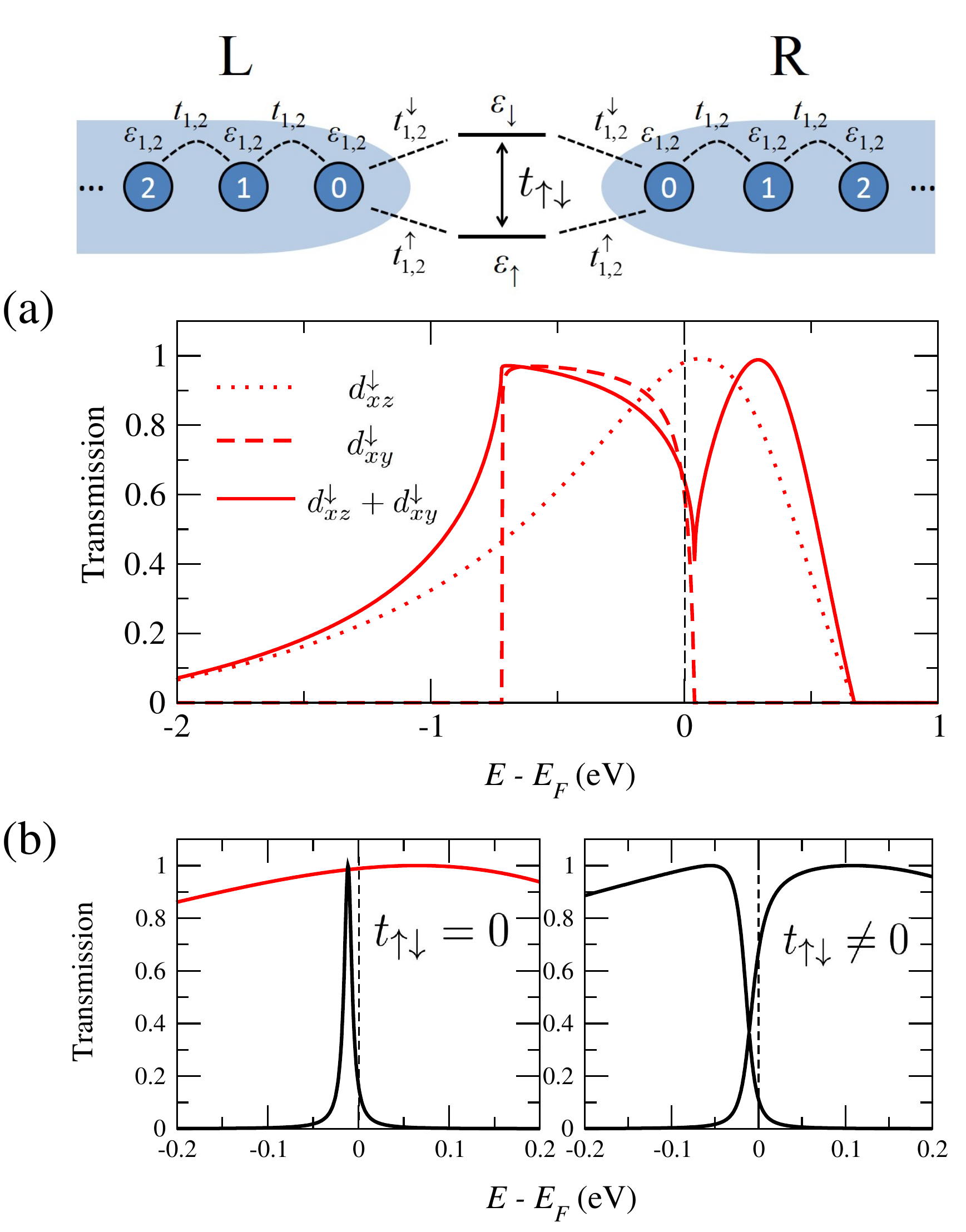}
	\caption{\label{tbmodel} 
		{
			Minimal tight-binding model explaining DFT results without SOC (a) and with SOC (b). 
			(a) A single (spin-down) level couples to two bands. The rather smooth transmissions due to the
			$d_{xz}^{\downarrow}$ band (dotted line) develops a pronounced dip right
			above $E_F$ if the coupling to the $d_{xy}^{\downarrow}$ band is switched on. 
			(b) Transmission eigenvalues mediated by two levels
			(each coupled to one band) without (left) and with (right) inter-level
			hopping, showing two independent or mixed channels, respectively. 
		}           
	}
\end{figure}

Based on the above arguments, we can understand main features of our
DFT results by setting up an appropriate tight-binding (TB) model with 
the following Hamiltonian:
\begin{equation}
\begin{aligned}
H=
\sum_{\alpha = \uparrow,\downarrow}\epsilon_{\alpha} \hat{c}^{\dagger}_{\alpha}\hat{c}_{\alpha}+(t_{\uparrow\downarrow}\hat{c}^{\dagger}_{\uparrow}\hat{c}_{\downarrow}+{\rm H.c.})+\\
\sum_{i=1}^{2}\sum_{j \in L,R}[\epsilon_i \hat{c}^{\dagger}_{i,j}\hat{c}_{i,j}+(t_i\hat{c}^{\dagger}_{i,j}\hat{c}_{i,j+1}+{\rm H.c.})]+\\
\sum_{\alpha = \uparrow,\downarrow}\sum_{i=1}^{2}[t_{i}^{\alpha}\hat{c}^{\dagger}_{\alpha}\hat{c}_{i,L0}+
t_{i}^{\alpha}\hat{c}^{\dagger}_{\alpha}\hat{c}_{i,R0}+ {\rm H.c.}].
\end{aligned}
\end{equation}
Here the first line considers HOMO spin-up and spin-down levels with the
hopping $t_{\uparrow\downarrow}$ between them, the second line describes the
left and right semi-infinite Ni chains with two bands, and the third line
refers to the coupling between the molecular levels and the two bands of the
chains. 
We extracted required TB parameters from \emph{ab initio} calculations by inspecting the band structures and by projecting the self-consistent Hamiltonian within the plane-wave basis onto atomic orbitals contained in the pseudo-potential files for each atom type. To double-check and refine our TB parameters, maximally localized Wannier functions were also used to reproduce electronic bands and transmissions of plane-wave DFT calculations. This was accomplished using $\textsc{Wannier90}$\cite{mostofi2008} code. A good agreement was found between Wannier TB and our TB Hamiltonian as can be seen in Fig. \ref{wannier} in the supplementary material.

We first model the transmission dip in the spin-down channel in the absence of
SOC. For this purpose we consider two bands, a wide $d^{\downarrow}_{xz}$ band
($\epsilon_1=-0.93$ eV, $t_1=0.8$ eV) and a narrow $d^{\downarrow}_{xy}$ band
($\epsilon_2=-0.34$ eV, $t_2=-0.19$ eV), which can couple to the HOMO
spin-down orbital at energy $\epsilon_{\downarrow}=-0.25$ eV (it is not seen in Fig. \ref{dft-trans}a due to strong hybridization with the $d^{\downarrow}_{xz}$ and $d^{\downarrow}_{xy}$). As seen in
Fig.~\ref{tbmodel}(a), when the HOMO couples only to $d_{xz}^{\downarrow}$
($t_1^{\downarrow}=-0.45$ eV) or to $d_{xy}^{\downarrow}$
($t_2^{\downarrow}=0.21$ eV), regular-shaped transmissions without a dip are
obtained.

When both couplings are taken into account however, a dip in
transmission develops right above the $d^{\downarrow}_{xy}$ band edge at $E-E_F=0.04$ eV, in agreement with the \textit{ab initio} results in
Fig.~\ref{dft-trans}(b). This dip can be seen as a result of destructive quantum
interference between two pathways as follows.  At energies $E-E_F<0.04$~eV the $d_{xy}^{\downarrow}$ Ni states form an  
additional conduction channel in the Ni chain, while for $E-E_F>0.04$~eV they contribute to an extra density of states (DOS) 
at the apex Ni atoms due to hybridization with the HOMO orbital. 
Those states will provide a second pathway for electron propagation:
Ni $d_{xz}^{\downarrow}\rightarrow$ HOMO spin-down $\rightarrow$ Ni-apex $d_{xy}^{\downarrow}\rightarrow$ HOMO spin-down
$\rightarrow$ Ni $d_{xz}^{\downarrow}$, in addition to the direct pathway: Ni $d_{xz}^{\downarrow}\rightarrow$
HOMO spin-down $\rightarrow$ Ni $d_{xz}^{\downarrow}$.
Since both pathways involve the same terminal $d_{xz}^{\downarrow}$ Ni band they will interfere (destructively) producing the
observed antiresonance in the transmission.    

We analyze now the case with SOC, aiming at explaining in particular the sharp
transmission feature just below the Fermi energy (see Fig.~\ref{dft-trans}(c)),
which is very different for the two magnetic orientations. Both HOMO spin-up
and spin-down molecular orbitals need to be included
($\epsilon_{\uparrow}=-0.015$~eV, $\epsilon_{\downarrow}=-0.25$~eV), which
couple to two Ni bands, $d^\downarrow_{yz}$ (channel 1)
and $d^\downarrow_{xz}$ (channel 2) ($\epsilon_{1,2}=-0.93$ eV, $t_{1,2}=0.8$ eV). 
Two other bands, $d^\downarrow_{xy}$ and $d^\downarrow_{x^2-y^2}$, are not relevant here and are disregarded for simplicity
(or can be considered as admixing in some minor proportion into two main channels mentioned above). 
The hopping
parameters are set to $t_1^{\uparrow}=0.05$~eV and
$t_2^{\downarrow}=-0.45$~eV, where the absolute value of $t_1^{\uparrow}$ is much smaller than  those of $t_2^{\downarrow}$ since it is purely due to SOC, while
$t_1^{\downarrow}=t_2^{\uparrow}=0$.  In the case of $\vec{M} \parallel \vec{x}$, as Eq. \ref{x-homo} shows, the two HOMO states are not mixed ($t_{\uparrow\downarrow}=0$) and
couple to the two independent Ni bands, which naturally yields two independent
conduction channels, see Fig.~\ref{tbmodel}(b).  
In the case of $\vec{M} \parallel \vec{z}$ (see Eq. \ref{z-homo})
a small inter-level hopping of
$t_{\uparrow\downarrow}=0.06$~eV should be introduced, which turns out to mix
the two channels and leads to their crossing, see Fig.~\ref{tbmodel}(b). This
simple model essentially reproduces our DFT results
(Fig.~\ref{T-eigenvalues}). 
Two key parameters introduced above, 
$t_1^{\uparrow}$ and $t_{\uparrow\downarrow}$, originate purely from SOC at the Ni/molecule
contacts and depend both on Ni SOC strength and on the HOMO composition (which can be inferred from
Eqs.\ref{homo-wave},\ref{x-homo},\ref{z-homo}).
Unlike other parameters (which could be extracted from the DFT Hamiltonian as discussed above), 
$t_1^{\uparrow}$ and $t_{\uparrow\downarrow}$ were determined by fitting the width (controlled by $t_1^{\uparrow}$) and the shape (controlled by $t_{\uparrow\downarrow}$) of model transmissions to DFT curves in Fig.~\ref{T-eigenvalues}.

Finally, it should be emphasized that the tilting of the BDT molecule in the
$yz$ plane (see Fig.~\ref{dft-trans}) is crucial for observing both the
transmission dip and the $\vec{M}$-dependent conduction channel crossing,
discussed in Fig.~\ref{tbmodel}. Due to
the tilting, the HOMO will also hybridize with Ni $d_{xy}$ states, in addition
to $d_{xz}$. This is
essential for providing (i) an additional pathway for electron propagation,
which causes the transmission dip, and (ii) mixing of HOMO spin-up and
spin-down states for the $\vec{M} \parallel \vec{z}$ orientation, which causes
the crossing of the two channels. The latter can be clearly seen from
Eqs.~\ref{homo-wave} and \ref{z-homo}, since the two HOMO orbitals remain
completely decoupled, if the coefficient $B_\uparrow=0$, which is the case for
a straight molecular orientation.

In conclusion, using fully relativistic DFT calculations, we find a very high
and energy-dependent AMR at the Fermi energy in Ni-BDT-Ni molecular
junctions. It stems from the SOC term, which opens a new conduction channel
via spin-flip processes at the ferromagnet-molecule interface. In the absence
of SOC, the channel was fully blocked due to the symmetry mismatch between the
involved HOMO orbital and the Ni electrode states. Importantly, this
HOMO-related conductance change is very sensitive to the magnetization
direction, resulting in a giant AMR right at the Fermi level. Moreover, a
significant AMR of about 95\% is found just above $E_F$ due to quantum
interference effects. A simple tight-binding model explains the main features
of our \textit{ab initio} results. Since the geometry of a molecular junction
depends on electrode separation, the AMR can be tuned by mechanical control,
as shown in Fig. \ref{strain} in the supplementary material.  We expect that the proposed mechanism, based on orbital symmetry reasonings, is generally at work in metal-molecule-metal junctions and
explains the high AMR values reported recently
\cite{li2015giant,yang2019tunable}. Our study reveals the general principles
that lead to an enhanced AMR in molecule-based spintronic devices.

{\bf Acknowledgments}: During this research project, D.L.\ was supported by the Alexander von
Humboldt Foundation through a Fellowship for Postdoctoral
Researchers. Furthermore, D.L.\ and F.P.\ acknowledge financial support from
the Collaborative Research Center (SFB) 767 of the German Research
Foundation (DFG).  Part of the numerical modeling was performed using the
computational resources of the bwHPC program, namely the bwUniCluster and the
JUSTUS HPC facility.
\\

\section*{Appendix A: Geometry optimization}
The geometry optimization of molecular junctions was performed in a supercell
containing a single BDT molecule and two four-atom Ni pyramids attached to a
Ni(111)-4 $\times$ 4 periodic slab with 16 atoms per layer and with five and
four layers on left and right sides, respectively. During the ionic relaxation
the three outermost Ni layers on both sides were kept fixed at bulk
structures, while the molecule and the other slab layers were allowed to relax
until atomic forces fell below 10$^{-3}$ Ry/Bohr. The geometry optimization
was performed using a 2 $\times$ 2 $\times$ 1 $\mathbf{k}$-point mesh. A
plane-wave basis was employed with an energy cutoff of 30 and 300 Ry for
wavefunctions and the charge density, respectively.

\section*{Appendix B: Transmission calculations of junctions}
\textit{Ab initio} transport properties including SOC were evaluated with the
$\textsc{pwcond}$ code \cite{Alexander2004}. Here the Ni(111) crystalline
electrodes were replaced by semi-infinite atomic chains. The SOC effect was
taken into account via fully relativistic pseudopotentials
\cite{Andrea2006}. The Hamiltonian is therefore a 2 $\times$ 2 matrix in
spin space, and the non-diagonal matrix elements arise from SOC. All the
calculations were done in the non-collinear mode with the specific
magnetization direction aligned along $\mathbf{z}$ or $\mathbf{x}$ axes. Separate
calculations were performed for the leads (complex band structure
calculations) and scattering regions, which were combined using the
wave-function matching technique. The self-consistency criterion in the DFT
calculations was set to 10$^{-8}$~Ry in order to obtain well-converged charge
and spin magnetization densities.

Our TB parameters, including on-site energies and hopping integrals, were extracted from \textit{ab initio} QE calculations by projecting the self-consistent Hamiltonian onto the basis of atomic wave functions provided by pseudo-potential files. This procedure is rather similar to the one used for calculating the projected density of states (PDOS). Only nearest-neighbor hopping is considered in Ni wires. To calculate couplings of molecular orbitals to Ni electrodes we first diagonalize the molecular Hamiltonian -- the Hamiltonian matrix restricted to the molecule atomic orbitals -- and then rotates the molecule coupling matrices from the atomic basis to molecular orbitals. We keep then only HOMO orbital and its coupling constants to the contact Ni atoms.

\begin{figure*}
	\includegraphics[scale=0.55]{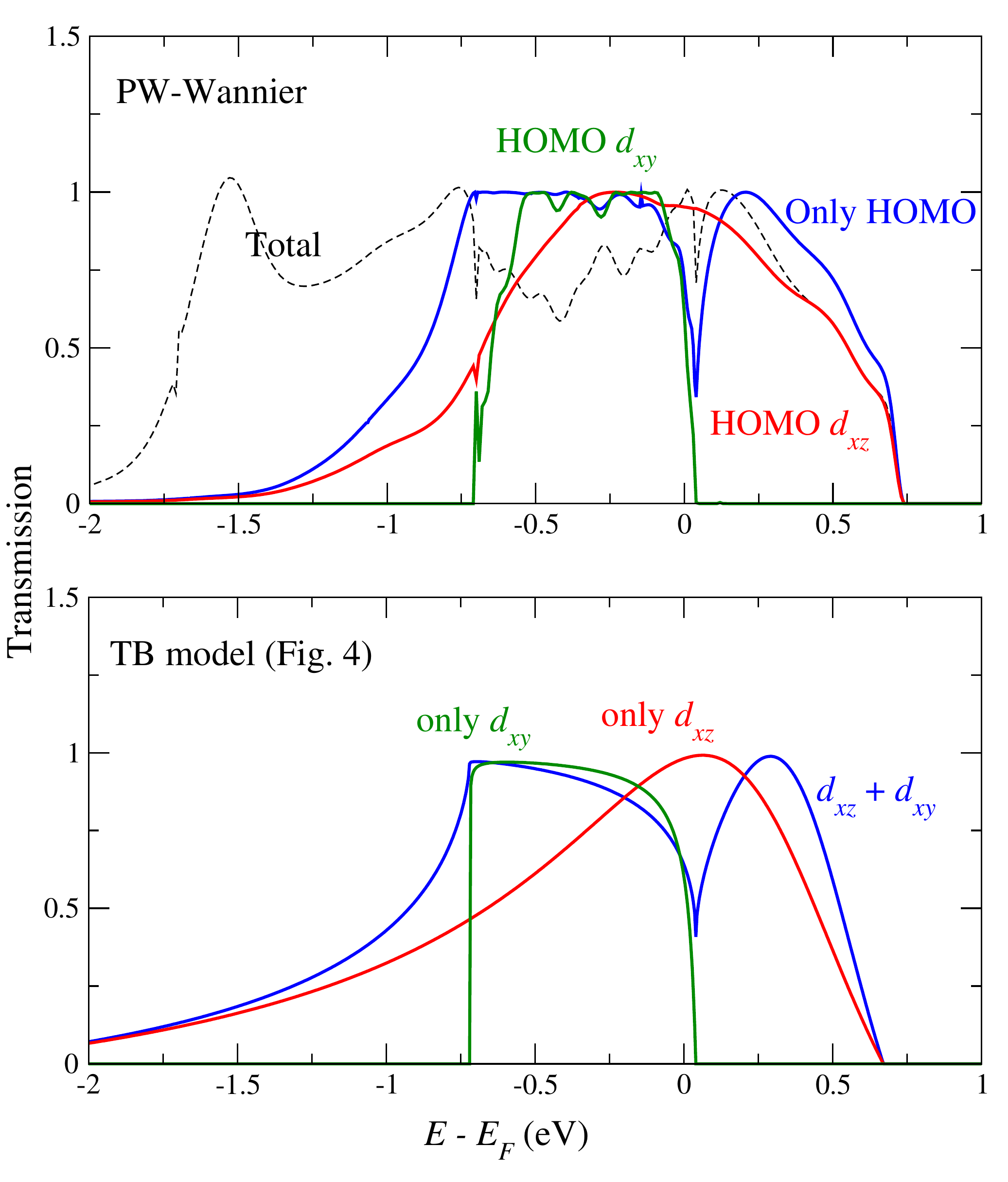}
	\caption{\label{wannier} Comparison of transmission functions calculated with Wannier functions Hamiltonian (top) and with the minimal TB model presented in the main text (bottom).}
\end{figure*}

To validate our minimal TB model and adjust TB parameters, we compare it with transmission calculations based on Wannier functions (WFs) which represent also a localized basis set (Fig. S2).  The WFs and the Hamiltonian were constructed from DFT Hamiltonian using $\textsc{Wannier90}$\cite{mostofi2008} code. Since WFs represent a complete basis set by construction (in an energy window of interest), the total spin down WFs transmission (top panel, dashed line) is in a very good agreement with the DFT curve in Fig. 2(b) of the main text.To compare directly with the TB model, we calculated the transmission only through the HOMO by setting artificially to zero coupling parameters for all other molecular orbitals (top panel, blue line). Finally, this HOMO transmission is further decomposed into $d_{xy}$- and $d_{xz}$-like components by keeping the coupling of HOMO to only $d_{xy}$ or $d_{xz}$ Ni bands, respectively. Comparing two panels of Fig. \ref{wannier} we can deduce that  our minimal TB Hamiltonian (with parameters presented in the main text) gives transmission curves (bottom panel, the same as in Fig. \ref{tbmodel}(a) of the main text) which agree rather well to ``exact" Wannier-based ones (top panel).

\section*{Appendix C: Controlling AMR via a mechanical strain}

\begin{figure*}
	\includegraphics[scale=0.7]{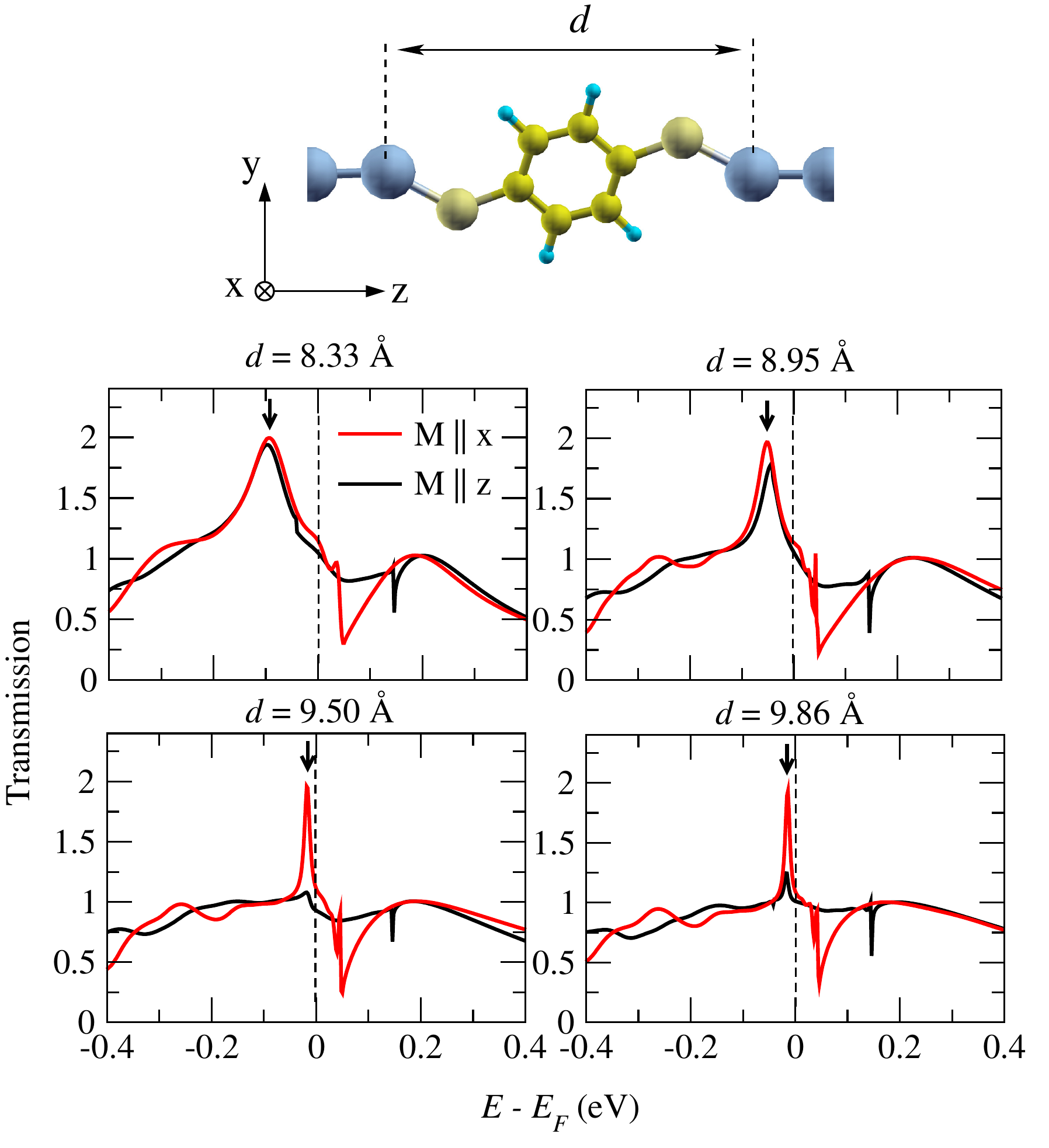}
	\caption{\label{strain} Controlling AMR via electrode separation. The degree of MAMR just
		below $E_{{F}}$ (marked with a downward-pointing arrow) can be tuned by
		stretching of the molecular junction due to a competition between
		hybridization and SOC effects.}
\end{figure*} 

\bibliographystyle{apsrev}
\bibliography{manuscript}

\end{document}